\documentclass[preprint,aps,showkeys,showpacs]{revtex4}
\usepackage[english]{babel}
\usepackage{graphicx}
\usepackage{graphics}
\usepackage{amsmath}
\usepackage{dcolumn}
\usepackage{amssymb}
\usepackage{bm}
\usepackage[latin1]{inputenc}

\begin{document}

\title{Abelian geometric phase for a Dirac neutral particle in a Lorentz
symmetry violation environment}
\author{K. Bakke}
\email{kbakke@fisica.ufpb.br}
\affiliation{Departamento de F\'isica, Universidade Federal da Para\'iba, Caixa Postal
5008, 58051-970, Jo\~ao Pessoa, PB, Brazil.}
\author{H. Belich}
\affiliation{Departamento de Física e Química, Universidade Federal do Espírito Santo,
Av. Fernando Ferrari, 514, Goiabeiras, 29060-900, Vitória, ES, Brazil.}

\begin{abstract}
We introduce a new term into the Dirac equation based on the Lorentz
symmetry violation background in order to make a theoretical description of
the relativistic quantum dynamics of a spin-half neutral particle, where the
wave function of the neutral particle acquires a relativistic Abelian
quantum phase given by the interaction between a fixed time-like $4$-vector
background and crossed electric and magnetic fields, which is analogous to
the geometric phase obtained by Wei \textit{et al} [H. Wei, R. Han and X.
Wei, Phys. Rev. Lett. \textbf{75}, 2071 (1995)] for a spinless neutral
particle with an induced electric dipole moment. We also discuss the flux
dependence of energy levels of bound states analogous to the Aharonov-Bohm
effect for bound states.
\end{abstract}

\maketitle

\section{Introduction}

In recent years, the appearance of topological or geometric phases has
attracted a great deal of interest in several physical systems such as in
condensed matter \cite{bohm}, in cosmological backgrounds \cite{bf7} and for
scalar quantum particles in spacetimes generated by defects \cite{fur4}. The
Aharonov-Bohm effect \cite{ab} is the best famous topological effect which
has been studied in scattering problems, and in the flux dependence of the
energy levels of bound states \cite{pesk}. Another interesting study of
topological effects in quantum systems is the dual effect of the
Aharonov-Bohm effect made by Dowling \textit{et al} \cite{dab}, and Furtado 
\textit{et al} \cite{dab2}. A well-known quantum effect related to the
appearance of quantum phases is the Aharonov-Casher effect \cite{ac}, which
can be considered as the reciprocal effect of the Aharonov-Bohm effect. More
discussions about topological quantum effects can be found in the literature
as the equivalence between the Aharonov-Bohm and the Aharonov-Casher effects 
\cite{ac2,ac3}, the topological nature of the Aharonov-Casher effect \cite%
{ac4,ac5}, the nonlocality and the nondispersivity \cite%
{ac6,ac7,disp,disp2,disp3}. Other studies based on the Aharonov-Casher
system have been done in the Landau quantization \cite{er,bf6,bf16,bbs},
bound states analogous to a quantum dot induced by noninertial effects \cite%
{b2}, the appearance of geometric phases in the noncommutative quantum
mechanics \cite{fur1}, and in the presence of topological defects \cite{bf3}.

The dual effect of the Aharonov-Casher effect has been proposed by He and
McKellar \cite{hm}, and Wilkens \cite{w}, which is known as the
He-McKellar-Wilkens effect. Involving a non-constant electric dipole moment,
Wei \textit{et al} \cite{whw} proposed a new system where the wave function
of a neutral particle acquires a quantum phase when the induced electric
dipole moment interacts with a configuration of crossed electric and
magnetic fields. Several discussions have been done about the topological
nature of the quantum phases for electric dipoles \cite%
{hag,whw2,aud,spa,spa1,spa2,spa3}, and experiments have also been proposed
with the purpose of producing the He-McKellar-Wilkens setup \cite%
{tka1,tka2,tka3,tka4,tka}. The solid state analogue of the
He-McKellar-Wilkens quantum phase has been discussed in \cite{fur2}.
Furthermore, bound states for a neutral particle with an induced electric
dipole moment have been studied in the Landau quantization \cite{fur3}, and
in the presence of an azimuthal magnetic field and in the presence of a
disclination \cite{b3}.

Recently, general discussions about quantum effects involving magnetic and
electric dipole moments have been done in the nonrelativistic context, and
in the relativistic context of quantum mechanics \cite%
{anan,anan2,sil,bjd,salpeter}. For instance, the relativistic quantum
dynamics of a neutral particle with permanent magnetic dipole moment can be
described by introducing a nonminimal coupling into the Dirac equation given
by $i\gamma^{\mu}\,\partial_{\mu}\rightarrow\,i\gamma^{\mu}\,\partial_{\mu}+%
\frac{\mu}{2}\,\Sigma^{\mu\nu}\,F_{\mu\nu}$ \cite{anan,anan2,sil,bjd}, where 
$\mu$ is the permanent magnetic dipole moment of the neutral particle, $%
F_{\mu\nu}$ is the electromagnetic tensor and $\Sigma^{\mu\nu}=\frac{i}{2}%
\left[\gamma^{\mu},\gamma^{\nu}\right]$, with $\gamma^{\mu}$ being the Dirac
matrices. For a neutral particle with permanent electric dipole moment, the
relativistic behavior has been introduced in the Dirac equation by Salpeter 
\cite{salpeter}. In recent works \cite{anan,anan2,sil}, the relativistic
behaviour of a neutral particle with permanent electric dipole moment can be
described by the dual transformation $\mu\leftrightarrow d$, $\vec{E}%
\leftrightarrow -\vec{B}$ and $\vec{B}\leftrightarrow\vec{E}$ \cite{sil},
where the nonminimal coupling becomes $i\gamma^{\mu}\,\partial_{\mu}%
\rightarrow\,i\gamma^{\mu}\,\partial_{\mu}-i\frac{d}{2}\,\Sigma^{\mu\nu}\,%
\gamma^{5}\,F_{\mu\nu}$ \cite{salpeter,anan,anan2,sil}, with $d$ being the
permanent electric dipole moment of the neutral particle. Other studies of
the quantum dynamics of a neutral particle with a magnetic dipole moment
have been done based on the violation of the Lorentz symmetry \cite%
{belich,belich2,belich3,bbs2}. In Refs. \cite{belich,belich2,belich3,bbs2},
the quantum dynamics of a neutral particle is described by introducing a
nonminimal coupling into the Dirac equation given by $i\gamma^{\mu}\,%
\partial_{\mu}\rightarrow\,i\gamma^{\mu}\,\partial_{\mu}-g\,v^{\nu}\,%
\widetilde{F}_{\mu\nu}\,\gamma^{\mu}$, where $v^{\nu}$ is a fixed $4$-vector
which acts as the background which breaks the Lorentz symmetry, $\widetilde{F%
}_{\mu\nu}$ corresponds to the dual electromagnetic tensor, and $g$ is a
coupling constant. One should note that, for relativistic models, the study
of symmetry breaking can be extended by considering a background given by a $%
4$-vector field that breaks the symmetry $\mathcal{SO}\left( 1,3\right) $
instead of the symmetry $\mathcal{SO}\left(3\right)$. This is known in the
literature as the spontaneous violation of the Lorentz symmetry \cite%
{extra3,extra1,extra2}. The spontaneous violation of Lorentz symmetry was
first proposed in 1989 by Kostelecky and Samuel \cite{extra3} indicating the
possibility that the spontaneous violation of symmetry by a scalar field
could be extended to the string field theory. The consequence of this
extension is a spontaneous breaking of the Lorentz symmetry as in the the
electroweak theory. In the electroweak theory, a scalar field acquires a
nonzero vacuum expectation value that yields mass to gauge bosons (Higgs
Mechanism). In a similar way, one can find in the string theory that this
scalar field can be extended to a tensor field. At present days, theories
involving the Lorentz symmetry breaking are part of the proposal of the
Extended Standard Model \cite{col} as a possible extension of the minimal
Standard Model of the fundamental interactions, where the violation of the
Lorentz symmetry is implemented in the fermion section of the Extended
Standard Model by two CPT-odd terms, that is, $v_{\mu }\overline{\psi}
\gamma ^{\mu }\psi $ and $b_{\mu }\overline{\psi }\gamma _{5}\gamma
^{\mu}\psi $ (where $v_{\mu }$ and $b_{\mu }$ correspond to the
Lorentz-violating vector field backgrounds \cite%
{belich,belic,belich2,belich3}). Furthermore, the modified Dirac theory has
already been examined in \cite{Hamilton}, and the spectrum of energy of the
hydrogen atom has been discussed in the nonrelativistic limit of the
modified Dirac theory in \cite{Manojr,Nonmini}. We can also find in the
literature an extensive amount of work looking at the violation of the
Lorentz symmetry, and numerous experimental bounds were estimated \cite%
{extra2}. However, is the introduction of this new nonminimal coupling $%
i\gamma ^{\mu}\,\partial _{\mu }\rightarrow \,i\gamma ^{\mu }\,\partial
_{\mu}-g\,v^{\nu }\,\widetilde{F}_{\mu \nu }\,\gamma^{\mu}$ consistent with
studies of the Lorentz violation symmetry? We will adopt the following point
of view: a nonminimal coupling is conceived as corrections of a dynamical
field, thus, the theory goes on to describe a new behavior, when the
physical system begin to access a new energy scale. Such coupling with the
background can present relevant information about the low energy regime of a
fundamental theory. Hence, if we have a fundamental theory with vector
fields (or tensor) which violate the Lorentz symmetry, by assuming
non-trivial expected values in vacuum, we propose that the fermion sector
should feels this background by a nonminimal coupling with a vector
background \cite{belich,belich2,belich3}. Specifically, in our case, such
coupling implies that the background can generate a Aharonov-Casher-type
phase to a neutral particle.

In this paper, based on the introduction of a nonminimal coupling into the
Dirac equation to describe the quantum dynamics of a neutral particle in the
Lorentz symmetry violation background \cite{belich,belich2,belich3,bbs2}, we
suggest the introduction of a new term into the Dirac equation based on the
Lorentz symmetry violation background to describe the relativistic quantum
dynamics of a spin-half neutral particle, where the wave function of the
neutral particle acquires a relativistic Abelian quantum phase analogous to
the geometric phase obtained by Wei \textit{et al} \cite{whw} for an induced
electric dipole moment. We show that the relativistic geometric phase
differs from the relativistic Aharonov-Casher geometric phase obtained in
Ref. \cite{bbs2} based on the Lorentz symmetry violation background, since
the relativistic geometric phase obtained in this work is given by the
interaction between a fixed time-like $4$-vector background and crossed
electric and magnetic fields. At the end, we also show that the energy
levels for bound states of a spin-half neutral particle confined to moving
between two coaxial cylinders are flux dependent. The structure of this
paper is: in section II, we introduce a new term in the Dirac equation based
on the Lorentz symmetry violation background, and discuss the relativistic
analogue of the geometric quantum phase obtained by Wei \textit{et al} \cite%
{whw} for a spin-half neutral particle, which is given by the interaction
between a fixed time-like $4$-vector background and crossed electric and
magnetic fields. We also discuss the flux dependence of the relativistic
energy levels for bound states; in section III, we present our conclusions.

\section{Relativistic Abelian geometric phase based on the Lorentz symmetry
breaking}

In this section, we suggest the introduction of a new term into the Dirac
equation based on the Lorentz symmetry violation background to describe a
relativistic quantum system, where the wave function of the neutral particle
acquires a relativistic quantum phase analogous to the geometric phase
obtained by Wei \textit{et al} \cite{whw} in the rest frame of the neutral
particle. Furthermore, we discuss the nonrelativistic limit of the Dirac
equation, and obtain a geometric quantum phase for a spin-half particle
analogous to that one obtained by Wei \textit{et al} \cite{whw}. At the end
of the section, we discuss the flux dependence of the energy levels for a
relativistic neutral particle confined to moving between two coaxial
cylinders. In that way, we introduce this new term in the Dirac equation as
follows 
\begin{eqnarray}
i\gamma^{\mu}\,\partial_{\mu}\rightarrow i\gamma^{\mu}\,\partial_{\mu}-\frac{%
g}{2}\,\eta^{\alpha\beta}\,F_{\mu\alpha}\,F_{\beta\nu}\,\gamma^{\mu}\,b_{%
\lambda}\gamma^{\lambda}\,\gamma^{\nu}-\frac{\nu}{2}\,\eta^{\alpha\beta}%
\,F_{\mu\alpha}\,F_{\beta\nu}\,\gamma^{\mu}\,\gamma^{\nu},  \label{1}
\end{eqnarray}
where $g$ and $\nu$ are constants, $b_{\lambda}$ corresponds to the fixed $4$%
-vector that breaks the Lorentz symmetry, $F_{\mu\nu}$ is the
electromagnetic field tensor whose components are $F_{0i}=-F_{i0}=-E_{i}$, $%
F_{ij}=-F_{ji}=\epsilon_{ijk}\,B^{k}$, and the matrices $\gamma^{\mu}$ are
the Dirac matrices given in the Minkowski spacetime \cite{bjd,greiner}, 
\textit{i.e.}, 
\begin{eqnarray}
\gamma^{0}=\hat{\beta}=\left( 
\begin{array}{cc}
I & 0 \\ 
0 & -I \\ 
& 
\end{array}%
\right);\,\,\, \gamma^{i}=\hat{\beta}\,\hat{\alpha}^{i}=\left( 
\begin{array}{cc}
0 & \sigma^{i} \\ 
-\sigma^{i} & 0 \\ 
& 
\end{array}%
\right);\,\,\,\Sigma^{i}=\left( 
\begin{array}{cc}
\sigma^{i} & 0 \\ 
0 & \sigma^{i} \\ 
& 
\end{array}%
\right),  \label{2}
\end{eqnarray}
with $I$ being the $2\times2$ identity matrix, $\vec{\Sigma}$ being the spin
vector, and $\sigma^{i}$ being the Pauli matrices. The Pauli matrices
satisfy the relation $\left(\sigma^{i}\,\sigma^{j}+\sigma^{j}\,\sigma^{i}%
\right)=2\,\eta^{ij}$, where $\eta^{\mu\nu}=\mathrm{diag}\left(-\,+\,+\,+%
\right)$ is the Minkowski tensor. By taking the Lorentz symmetry violation
background given by a time-like vector $b^{\lambda}=\left(b^{0},0,0,0\right)$
in such a way that we can consider $gb^{0}=\nu$ \footnote{%
We have chosen to discuss the case where $gb^{0}=\nu$ because we intend to
obtain a nonrelativistic system analogous to that one worked in Ref. \cite%
{whw} from the Lorentz symmetry violation background. In the case $%
gb^{0}\neq\nu$, we can see that we can obtain the relativistic Abelian
geometric phase, but the nonrelativistic system has two different coupling
constants, thus, it differs from that one of Ref. \cite{whw}. In this way,
it makes sense to take $\nu=0$, and obtain just the geometric phase without
discussing the influence of local terms in this quantum dynamics.}, and
using the definitions of $F_{\mu\nu}$ and $\Sigma^{\mu\nu}$ given earlier,
we can develop the nonminimal coupling term of the expression (\ref{1}) and
obtain $\frac{\nu}{2}\,\eta^{\alpha\beta}\,F_{\mu\alpha}\,F_{\beta\nu}\,%
\gamma^{\mu}\left(\gamma^{0}+I\right)\,\gamma^{\nu}=\nu\,\hat{\beta}\,\vec{%
\alpha}\cdot\left(\vec{E}\times\vec{B}\right)-2\nu\,B^{2}\,\hat{\beta}%
-\nu\,\left(E^{2}-B^{2}\right)$. With direct calculations, we can verify
that the new term introduced in the equation (\ref{1}) is Hermitian, thus,
we have a toy model to study relativistic geometric quantum phases for a
spin-half neutral particle analogous to a neutral particle with an induced
electric dipole moment based on the background where the Lorentz symmetry is
violated. We will show that we can make a study of relativistic quantum
mechanics in this background, where the relativistic analogue of the
geometric phase obtained by Wei \textit{et al} \cite{whw} can be obtained
and, in the nonrelativistic limit, we can obtain an analogous system to that
worked by Wei \textit{et al} \cite{whw} for a spinless neutral particle with
an induced electric dipole moment.

Now, let us consider the same the field configuration given in Ref. \cite%
{whw}. In this way, the spin-half neutral particle interacts with a
configuration of crossed electric and magnetic field given by 
\begin{eqnarray}
\vec{E}=\frac{\lambda}{\rho}\,\hat{\rho};\,\,\,\,\,\vec{B}=B_{0}\,\hat{z},
\label{8}
\end{eqnarray}
with $\lambda$ being a linear density of electric charges and $B_{0}$ being
a constant. In this system, it is convenient to work with the cylindrical
symmetry, where the line element of the Minkowski spacetime is written in
the form $ds^{2}=-dt^{2}+d\rho^{2}+\rho^{2}d\varphi^{2}+dz^{2}$. Since we
are working with curvilinear coordinates, it is convenient to treat the
Dirac spinors by using the mathematical formulation of spinor theory in
curved spacetime background \cite{weinberg}. In curved spacetime background,
the spinors are defined locally in the local reference frame of the
observers \cite{weinberg}. We can build the local reference frame for the
observers through a noncoordinate basis $\hat{\theta}^{a}=e^{a}_{\,\,\,\mu}%
\left(x\right)\,dx^{\mu}$, which components $e^{a}_{\,\,\,\mu}\left(x\right)$
satisfy the following relation $g_{\mu\nu}\left(x\right)=e^{a}_{\,\,\,\mu}%
\left(x\right)\,e^{b}_{\,\,\,\nu}\left(x\right)\,\eta_{ab}$ \cite{weinberg}.
The indices $(a,b,c=0,1,2,3)$ indicate the local reference frame of the
observers and the indices $\left(\mu,\nu=t,\rho,\varphi,z\right)$ indicate
the spacetime indices. The components of the noncoordinate basis $%
e^{a}_{\,\,\,\mu}\left(x\right)$ are called \textit{tetrads} and they form
the local reference frame of the observers. The tetrads have an inverse
defined by $dx^{\mu}=e^{\mu}_{\,\,\,a}\left(x\right)\,\hat{\theta}^{a}$,
where they are related by the expressions $e^{a}_{\,\,\,\mu}\left(x\right)%
\,e^{\mu}_{\,\,\,b}\left(x\right)=\delta^{a}_{\,\,\,b}$ and $%
e^{\mu}_{\,\,\,a}\left(x\right)\,e^{a}_{\,\,\,\nu}\left(x\right)=\delta^{%
\mu}_{\,\,\,\nu}$. Moreover, in curved spacetime background, the partial
derivative $\partial_{\mu}$ must be changed by the covariant derivative of a
spinor \cite{weinberg,naka} given by $\nabla_{\mu}=\partial_{\mu}+\Gamma_{%
\mu}\left(x\right)$, with $\Gamma_{\mu}\left(x\right)=\frac{i}{4}%
\,\omega_{\mu ab}\left(x\right)\,\Sigma^{ab}$ being the spinorial connection 
\cite{weinberg,naka}. Choosing $\hat{\theta}^{0}=dt$, $\hat{\theta}^{1}=d\rho
$, $\hat{\theta}^{2}=\rho\,d\varphi$ and $\hat{\theta}^{3}=dz$, we can
obtain the non-null components of $1$-form connection $\omega_{\mu
ab}\left(x\right)$ by solving the Maurer-Cartan structure equations $d\hat{%
\theta}^{a}+\omega^{a}_{\,\,\,b}\wedge\hat{\theta}^{b}=0$ \cite{naka}, with $%
\omega^{a}_{\,\,\,b}=\omega_{\mu\,\,\,b}^{\,\,\,a}\left(x\right)dx^{\mu}$.
Direct calculations give us $\omega_{\varphi\,\,\,2}^{\,\,\,1}\left(x%
\right)=-\omega_{\varphi\,\,\,1}^{\,\,\,2}\left(x\right)=-1$ and,
consequently, $\gamma^{\mu}\,\Gamma_{\mu}=\frac{\gamma^{1}}{2\rho}$ \cite%
{bf6}. Hence, based on the nonminimal coupling (\ref{1}), the Dirac equation
becomes 
\begin{eqnarray}
i\frac{\partial\psi}{\partial t}=m\,\hat{\beta}\psi+\vec{\alpha}\cdot\left[%
\vec{p}-i\vec{\xi}+\nu\,\vec{E}\times\vec{B}\right]\psi-\nu\,B^{2}\,\psi-\nu%
\,\hat{\beta}\left(E^{2}-B^{2}\right)\psi,  \label{4}
\end{eqnarray}
where we have defined the vector $\vec{\xi}$ whose non-null components are $%
-i\xi_{k}=-ie^{\varphi}_{\,\,\,k}\left(x\right)\Gamma_{\varphi}=-\frac{%
\Sigma^{3}}{2\rho}\,\delta_{2k}$. Note that, given the field configuration (%
\ref{8}), the operator $\hat{\Pi}=\vec{p}-i\vec{\xi}+\nu\,\vec{E}\times\vec{B%
}$ is not a constant of motion \footnote{%
This can be checked through the equation \cite{greiner} $\frac{d\hat{\Pi}}{dt%
}=i\left[H_{D},\hat{\Pi}\right]=\nu\,\vec{\alpha}\times\vec{B}_{\mathrm{eff}%
}+\vec{\nabla}\left(\nu\,E^{2}\right)$, where $\frac{\partial\hat{\Pi}}{%
\partial t}=0$, $H_{D}$ is the Dirac Hamiltonian which corresponds to the
right-hand-side of the equation (\ref{4}), and $\vec{B}_{\mathrm{eff}}=\vec{%
\nabla}\times\left(\vec{E}\times\vec{B}\right)$ is the effective magnetic
field. With the field configuration considered in Eq. (\ref{12}), we have
that $\vec{B}_{\mathrm{eff}}=0$. Thus, we have that $\frac{d\hat{\Pi}}{dt}=%
\vec{\nabla}\left(\nu\,E^{2}\right)$.}, but there are no classical forces
acting on the dipole moment due to the term $\nu\,\vec{E}\times\vec{B}$,
which indicates that the relativistic quantum phase generated by the
presence of the term $\nu\,\vec{E}\times\vec{B}$ in the Dirac equation (\ref%
{4}) has a topological nature in the same sense of that discussed in Refs. 
\cite{ac,ac4,whw}.

Let us first discuss the nonrelativistic limit of the Dirac equation (\ref{4}%
), and show that we can obtain a nonrelativistic equation of motion for a
spin-half neutral particle analogous to the nonrelativistic equation of
motion for a spinless neutral particle with an induced electric dipole
moment worked by Wei \textit{et al.} in \cite{whw}. We can obtain the
nonrelativistic dynamics of the neutral particle when we extract the
temporal dependence of the wave function due to the rest energy \cite%
{bjd,greiner}, thus, we write the Dirac spinor in the form $%
\psi=e^{-imt}\,\left(\phi\,\,\chi\right)^{T}$, where $\phi$ and $\chi$ are
two-spinors. Substituting this solution into the Dirac equation (\ref{4}),
we obtain two coupled equations for $\phi$ and $\chi$, where the first
coupled equation is $i\frac{\partial\phi}{\partial t}+\nu\,E^{2}\,\phi=\left[%
\vec{\sigma}\cdot\left(\vec{p}-i\vec{\xi}\right)+\nu\vec{\sigma}\cdot\left(%
\vec{E}\times\vec{B}\right)\right]\chi$. By considering $\phi$ being the
``large'' component and $\chi$ being the ``small'' component, we can take $%
\left|2m\chi\right|>>\left|i\frac{\partial\chi}{\partial t}\right|$, $%
\left|2m\chi\right|>>\frac{\nu}{2}\left(E^{2}+B^{2}\right)\chi$, and write
the second coupled equation in the form: $\chi\approx\frac{1}{%
2\left(m+\nu\,B^{2}\right)}\left[\vec{\sigma}\cdot\left(\vec{p}-i\vec{\xi}%
\right)+\nu\vec{\sigma}\cdot\left(\vec{E}\times\vec{B}\right)\right]\phi$
without loss of generality. In that way, substituting $\chi$ of the second
coupled equation into the first coupled equation and neglecting terms of
order $m^{-2}$, we obtain the following second order differential equation 
\begin{eqnarray}
i\frac{\partial\phi}{\partial t}\approx\frac{1}{2\left(m+\nu\,B^{2}\right)}%
\left[\vec{p}-i\vec{\xi}+\nu\left(\vec{E}\times\vec{B}\right)\right]%
^{2}\phi-\nu\,E^{2}\,\phi+\frac{\nu}{2\left(m+\nu\,B^{2}\right)}\,\vec{\sigma%
}\cdot\vec{B}_{\mathrm{eff}}\phi,  \label{7}
\end{eqnarray}
which corresponds to the nonrelativistic limit of the Dirac equation (\ref{4}%
). Note that, up to the term $-i\vec{\xi}$ and the last term of the equation
(\ref{7}), we have the Schr\"odinger equation for a spinless neutral
particle with an induced electric dipole moment worked in \cite{whw}. The
last term of the nonrelativistic equation (\ref{7}) appears due to the
interaction between the spin and an effective magnetic field given by $\vec{B%
}_{\mathrm{eff}}=\vec{\nabla}\times\left(\vec{E}\times\vec{B}\right)$.
Taking the field configuration (\ref{8}), we can see that the effective
magnetic field is null $\left(B_{\mathrm{eff}}=0\right)$, and the
nonrelativistic equation (\ref{7}) becomes the Schr\"odinger equation
obtained in \cite{whw} (up to the term $-i\vec{\xi}$). Moreover, we obtain
in Eq. (\ref{7}) the phase shift $\Phi_{\mathrm{NR}}=\oint\nu\left(\vec{E}%
\times\vec{B}\right)\cdot d\vec{r}=2\pi\,\nu\lambda B_{0}$, which
corresponds to the Abelian geometric phase analogous to that obtained by Wei 
\textit{et al} \cite{whw}, but for a spin-half neutral particle, where we
have obtained in the Lorentz symmetry violation background. Hence, we have
shown that the nonrelativistic limit of the Dirac equation (\ref{4}) yields
both the equation of motion and the Abelian geometric phase analogous to
that obtained by Wei \textit{et al} \cite{whw}, but, in this case, for a
spin-half neutral particle in the Lorentz symmetry violation background.

From now on, we will discuss the appearance of a relativistic geometric
phase analogous to the geometric phase obtained by Wei \textit{et al} \cite%
{whw}, when we consider the introduction of the new term in the Dirac
equation (\ref{1}), and the field configuration given in (\ref{8}). To
obtain the relativistic quantum phase, we apply the Dirac phase factor
method \cite{dirac}, where the wave function of the neutral particle is
written in the form $\psi=e^{i\Phi}\,\psi_{0}$. Substituting this ansatz
into the Dirac equation (\ref{4}), we have that $\psi_{0}$ is the solution
of the equation 
\begin{eqnarray}
i\gamma^{0}\frac{\partial\psi_{0}}{\partial t}+i\gamma^{1}\,\left(\frac{%
\partial}{\partial\rho}+\frac{1}{2\rho}\right)\psi_{0}+i\frac{\gamma^{2}}{%
\rho}\frac{\partial\psi_{0}}{\partial\varphi}+i\gamma^{3}\frac{%
\partial\psi_{0}}{\partial z}+\nu\,E^{2}\,\psi_{0}-m\psi_{0}=0,  \label{10}
\end{eqnarray}
where the term proportional to $E^{2}$ is a local term and does not
contribute to the geometric phase \cite{anan,anan2,fur1}. We also have that
the term proportional to $B^{2}$ provides a contribution to the dynamical
phase given by $\Phi_{D}=-\nu\,\int_{0}^{\tau}\left(\hat{\beta}%
-I\right)\,B^{2}\,dt=-\nu\,B_{0}^{2}\,\tau\,\hat{\beta}-\nu B_{0}^{2}\tau$,
which is equivalent to that one given in \cite{whw}. The relativistic
geometric quantum phase acquired by the wave function of the neutral
particle is 
\begin{eqnarray}
\Phi=-\nu\oint\left(\vec{E}\times\vec{B}\right)_{\varphi}\,d\varphi=2\pi\,%
\nu\lambda B_{0}.  \label{11}
\end{eqnarray}

The relativistic quantum phase (\ref{11}) is obtained by the interaction
between a fixed time-like $4$-vector and a configuration of crossed electric
and magnetic fields given in (\ref{8}). As discussed earlier, there are no
classical forces acting on the dipole moment due to the presence of the term 
$\nu\,\vec{E}\times\vec{B}$. Thus, the relativistic quantum phase (\ref{11})
has a topological nature \cite{ac,ac4} in the same way of the
nonrelativistic case given in \cite{whw}. Moreover, we can also see that the
relativistic quantum phase is nondispersive \cite{disp,disp2,disp3}, that
is, it does not depend of the velocity of the neutral particle. In this
sense, we can see that the expression (\ref{11}) corresponds to the
relativistic analogue of the geometric quantum phase obtained by Wei \textit{%
et al} \cite{whw} in the Lorentz symmetry violation background. Furthermore,
We can easily see that the relativistic geometric phase (\ref{11}) differs
from the relativistic Aharonov-Casher geometric phase based on the Lorentz
symmetry violation background obtained in \cite{bbs2}, since the
relativistic geometric phase (\ref{11}) is provided by the interaction
between a time-like $4$-vector background $\left(\nu=g\,b^{0}\right)$ and a
field configuration given by crossed electric and magnetic fields. Thus, the
relativistic geometric phase (\ref{11}) brings us a new result in the
studies of geometric phases for neutral particle under the influence of the
Lorentz symmetry breaking. By comparing with Ref. \cite{whw}, we should note
that the nature of the geometric phase obtained by Wei \textit{et al} \cite%
{whw} is a consequence of the response to the particle with respect to
electric and magnetic field applied, therefore it depends on the
polarization which is a characteristic of the particle. However, the phase
shift analysed in this work is generated the zero component of the $4$%
-vector background $b^{0}$. This $4$-vector generates an anisotropy in the
spacetime, thus, if we change the particle, the response of the particle
changes, but the geometric phase produced by the $4$-vector background
remains unchanged. Thereby, if we want to analyse a neutral particle with
the settings in the electric and magnetic fields given in \cite{whw} plus
the Lorentz symmetry violation background in the nonrelativistic limit, we
have two independent contributions for the geometric phase. One should
expect that the coming phase from the presence of the $4$-vector background
being weaker than that generated by the polarization of the particle, but
there exists this contribution to the geometric phase. For the relativistic
Abelian geometric phase (\ref{11}), we can also expect that this phase is
quite small, thus, one should repeat the closed path given in (\ref{11})
many times in order to obtain a significant contribution to the geometric
phase from the $4$-vector background. For instance, let us suppose an
experimental ability to measure geometrical phases as small as $10^{-4}%
\mathrm{rad}$ \cite{cimmino}, thus, we can affirm that the theoretical phase
induced for a neutral particle can not be larger than this value, that is, $%
\nu \lambda B_{0}\,<\,10^{-4}\mathrm{rad}$. In this way, by taking the
values of the fields $\left|\vec{E}\right|\approx10^{7}\,\frac{\mathrm{V}}{%
\mathrm{m}}$, $B_{0}\approx 5\mathrm{T}$ $r_{0}=10^{-5}\mathrm{m}$ (which
correspond to usual values of electric fields and radius for 1D mesoscopic
rings \cite{Nitta}), and working with the natural units system $\hbar=c=1$
(wherein $1\mathrm{V}=11.7\,\mathrm{eV}$), we can estimate a upper bound for
the constant $\nu=g\,b^{0}$ given by $\nu <10^{-12} \left(\mathrm{eV}%
\right)^{-3}$.

At this moment, let us discuss the bound states which arise in the
relativistic quantum dynamics of the neutral particle described by the Dirac
equation (\ref{4}) when we restrict the neutral particle to move in a region
between two coaxial cylinders with impenetrable walls. Let us take the
solution of the Dirac equation (\ref{4}) in the form: $\psi=e^{-i\mathcal{E}%
t}\,\left(\upsilon\,\,\varsigma\right)^{T}$, where $\upsilon$ and $\zeta$
are two-spinors. Thus, substituting this solution into (\ref{4}), we obtain
two coupled equations for $\upsilon$ and $\varsigma$, where the first
coupled equation is 
\begin{eqnarray}
\left(\mathcal{E}-m+\frac{\nu\lambda^{2}}{\rho^{2}}\right)\upsilon=\left[%
-i\sigma^{1}\frac{\partial}{\partial\rho}-\frac{i\sigma^{1}}{2\rho}-\frac{%
i\sigma^{2}}{\rho}\frac{\partial}{\partial\varphi}-i\sigma^{3}\frac{\partial%
}{\partial z}-\frac{\Phi}{2\pi\rho}\,\sigma^{2}\right]\varsigma,  \label{12}
\end{eqnarray}
and the second coupled equation is 
\begin{eqnarray}
\left(\mathcal{E}+m-\frac{\nu\lambda^{2}}{\rho^{2}}\right)\varsigma=\left[%
-i\sigma^{1}\frac{\partial}{\partial\rho}-\frac{i\sigma^{1}}{2\rho}-\frac{%
i\sigma^{2}}{\rho}\frac{\partial}{\partial\varphi}-i\sigma^{3}\frac{\partial%
}{\partial z}-\frac{\Phi}{2\pi\rho}\,\sigma^{2}\right]\upsilon.  \label{13}
\end{eqnarray}
Thus, by eliminating $\varsigma$ in the equation (\ref{13}) and neglecting
the terms proportional to $m^{-2}$ and $\nu^{2}$ (we consider $\nu$ being a
small coupling constant), we obtain the following second order differential
equation 
\begin{eqnarray}
\left[\mathcal{E}^{2}-m^{2}+\frac{2m\nu\lambda^{2}}{\rho^{2}}\right]%
\upsilon&=&-\frac{\partial^{2}\upsilon}{\partial\rho^{2}}-\frac{1}{\rho}%
\frac{\partial\upsilon}{\partial\rho}+\frac{\upsilon}{4\rho^{2}}-\frac{1}{%
\rho^{2}}\frac{\partial^{2}\upsilon}{\partial\varphi^{2}}-\frac{%
\partial^{2}\upsilon}{\partial z^{2}}+\frac{i\sigma^{3}}{\rho^{2}}\frac{%
\partial\upsilon}{\partial\varphi}  \notag \\
[-2mm]  \label{14} \\[-2mm]
&+&\sigma^{3}\,\frac{\Phi}{2\pi\rho^{2}}\upsilon+2i\frac{\Phi}{2\pi\rho}%
\frac{\partial\upsilon}{\partial\varphi}+\left(\frac{\Phi}{2\pi}\right)^{2}%
\frac{\upsilon}{\rho^{2}}.  \notag
\end{eqnarray}

We can see in Eq. (\ref{14}) that $\upsilon$ is an eigenfunction of $%
\sigma^{3}$ whose eigenvalues are $s=\pm1$. In that way, we can split $%
\upsilon$ into $\upsilon=\left(\upsilon_{+}\,\,\upsilon_{-}\right)^{T}$
where $\sigma^{3}\upsilon_{+}=\upsilon_{+}$ and $\sigma^{3}\upsilon_{-}=-%
\upsilon_{-}$. Hence, in order to solve the second order differential
equation for both components $\upsilon_{+}$ and $\upsilon_{-}$, we write
these components in the compact form $\upsilon_{s}$, with $%
\sigma^{3}\upsilon_{s}=s\,\upsilon_{s}$. Thus, we can take the solution of
the second order differential equation (\ref{14}) in the form $%
\upsilon_{s}=C\,e^{i\left(l+\frac{1}{2}\right)\varphi}\,e^{ikz}\,R_{s}\left(%
\rho\right)$, where $l$ is a integer number, $k$ is a constant and $C$ is a
constant. Substituting $\upsilon_{s}$ into (\ref{14}), we obtain 
\begin{eqnarray}
\left[\frac{d^{2}}{d\rho^{2}}+\frac{1}{\rho}\frac{d}{d\rho}-\frac{%
\left(\zeta_{s}^{2}-\tau^{2}\right)}{\rho^{2}}+\kappa^{2}\right]%
R_{s}\left(\rho\right)=0,  \label{16}
\end{eqnarray}
with $\zeta_{s}=l+\frac{1}{2}\left(1-s\right)-\frac{\Phi}{2\pi}$, $%
\tau^{2}=2m\nu\lambda^{2}$ and $\kappa^{2}=\mathcal{E}^{2}-m^{2}-k^{2}$. We
have that the equation (\ref{16}) is the Bessel differential equation and
the general solution for (\ref{16}) is given by $R_{s}\left(\rho%
\right)=A_{l}\,J_{\eta}\left(\kappa\rho\right)+B_{l}\,N_{\eta}\left(\kappa%
\rho\right)$, where $\eta=\sqrt{\gamma_{s}^{2}-\tau^{2}}$, $%
J_{\eta}\left(\kappa\rho\right)$ and $N_{\eta}\left(\kappa\rho\right)$ are
the Bessel functions of the first and second kind, respectively. Now, we
consider the neutral particle is restricted to move in a region between two
coaxial cylindrical surfaces $\rho=\rho_{a}$ and $\rho=\rho_{b}$, where $%
\rho_{b}>\rho_{a}$. Thus, by considering the boundaries of these regions as
impenetrable, we require that the wave function satisfies the boundary
conditions: $R\left(\rho_{a}\right)=R\left(\rho_{b}\right)=0$. These
conditions provide us the following equation for the energy spectrum of the
neutral particle 
\begin{eqnarray}
J_{\eta}\left(\kappa\rho_{a}\right)\,N_{\eta}\left(\kappa\rho_{b}\right)-J_{%
\eta}\left(\kappa\rho_{b}\right)\,N_{\eta}\left(\kappa\rho_{a}\right)=0.
\label{18}
\end{eqnarray}

In order to obtain the energy spectrum explicitly, we consider a situation
in which $\kappa\rho_{a}>>1$ and $\kappa\rho_{b}>>1$. Then, we apply the
well-known asymptotic of the Hankel function, when $\eta$ is fixed, so 
\begin{eqnarray}  \label{19}
J_{\eta}\left(\kappa\rho_{a}\right)&\approx&\sqrt{\frac{2}{\pi\kappa\rho_{a}}%
}\left[\cos\left(\kappa\rho_{a}-\frac{\eta\pi}{2}-\frac{\pi}{4}\right)-\frac{%
\left(4\eta^{2}-1\right)}{8\kappa\rho_{a}}\sin\left(\kappa\rho_{a}-\frac{%
\eta\pi}{2}-\frac{\pi}{4}\right)\right];  \notag \\
\\
N_{\eta}\left(\kappa\rho_{a}\right)&\approx&\sqrt{\frac{2}{\pi\kappa\rho_{a}}%
}\left[\sin\left(\kappa\rho_{a}-\frac{\eta\pi}{2}-\frac{\pi}{4}\right)+\frac{%
\left(4\eta^{2}-1\right)}{8\kappa\rho_{a}}\cos\left(\kappa\rho_{a}-\frac{%
\eta\pi}{2}-\frac{\pi}{4}\right)\right] ,  \notag
\end{eqnarray}
and by interchanging $\rho_{a}$ for $\rho_{b}$, we obtain similar
expressions for $J_{\eta}\left(\kappa\rho_{b}\right)$ and $%
N_{\eta}\left(\kappa\rho_{b}\right)$. Thus, substituting the expressions (%
\ref{19}) and similar expressions for $J_{\eta}\left(\kappa\rho_{b}\right)$
and $N_{\eta}\left(\kappa\rho_{b}\right)$ into (\ref{18}), we obtain $%
\kappa^{2}\approx\left(\frac{n\pi}{\rho_{b}-\rho_{a}}\right)^{2}+\frac{%
\left(4\eta^{2}-1\right)}{4\rho_{a}\rho_{b}}$, with $n=0,1,2,3\ldots$ Hence,
by using the definition of the parameters $\zeta_{s},\kappa,\tau$ given
earlier, the energy levels for this system are 
\begin{eqnarray}
\mathcal{E}_{n,\,l}^{2}&\approx& m^{2}+k^{2}+\frac{\left(n\pi\right)^{2}}{%
\left(\rho_{b}-\rho_{a}\right)^{2}}+\frac{\left[l+\frac{1}{2}%
\left(1-s\right)-\frac{\Phi}{2\pi}\right]^{2}}{\rho_{a}\rho_{b}}-\frac{%
2m\nu\lambda^{2}}{\rho_{a}\rho_{b}}-\frac{1}{4\rho_{a}\rho_{b}},  \label{21}
\end{eqnarray}

The expression (\ref{21}) corresponds to the relativistic energy levels for
bound states when the neutral particle is restrict to move in a confined
region between two coaxial cylinders surfaces with impenetrable walls. These
bound states depend on the relativistic geometric quantum phase $\Phi$ given
in the expression (\ref{11}) with periodicity $\phi_{0}=2\pi$. Note that the
dependence of the energy levels (\ref{21}) on the relativistic geometric
phase $\Phi$ provide us an analogous effect to the Aharonov-Bohm effect for
bound states \cite{pesk}. To obtain the nonrelativistic expression for the
energy levels, we must apply the Taylor expansion in the expression (\ref{21}%
), and obtain 
\begin{eqnarray}
\mathcal{E}_{n,\,l}&\approx& m+\frac{1}{2m}\frac{\left(n\pi\right)^{2}}{%
\left(\rho_{b}-\rho_{a}\right)^{2}}+\frac{1}{2m}\frac{\left[l+\frac{1}{2}%
\left(1-s\right)-\frac{\Phi}{2\pi}\right]^{2}}{\rho_{a}\rho_{b}}-\frac{%
\nu\lambda^{2}}{\rho_{a}\rho_{b}}-\frac{1}{8M\rho_{a}\rho_{b}}+\frac{k^{2}}{%
2m},  \label{22}
\end{eqnarray}
where we must remember that $m$ is the rest mass of neutral particle and the
remaining terms of the equation (\ref{22}) correspond to the nonrelativistic
energy levels for bound states. Note that the energy levels for the bound
states depend on the quantum flux $\Phi$ as obtained in \cite{whw}. In order
to compare with the nonrelativistic energy levels of the bound states given
in \cite{whw} (where it is considered that the neutral particle moves in a
ring of radius $R$), from the expression (\ref{21}), we can see that when $%
\rho_{b}\rightarrow\rho_{a}\Rightarrow\mathcal{E}\rightarrow\infty$. Thus,
to obtain the limit where $\mathcal{E}=\mathrm{const}$ as $%
\rho_{b}\rightarrow\rho_{a}$, we must introduce an attractive potential in
the region $\rho_{a}\,<\,\rho\,<\,\rho_{b}$ in order to compensate the
energy increase of the radial modes in this limit \cite{val}. In that way,
the energy levels (\ref{21}) becomes 
\begin{eqnarray}
\mathcal{E}_{l}^{2}\approx m^{2}+k^{2}+\frac{\left[l+\frac{1}{2}%
\left(1-s\right)-\frac{\Phi}{2\pi}\right]^{2}}{\rho_{a}^{2}}-\frac{%
2m\nu\lambda^{2}}{\rho_{a}^{2}}-\frac{1}{4\rho_{a}^{2}}.  \label{23}
\end{eqnarray}
Taking the nonrelativistic limit of the expression (\ref{23}) through the
Taylor expansion, we obtain 
\begin{eqnarray}
\mathcal{E}_{l}\approx m+\frac{1}{2m}\frac{\left[l+\frac{1}{2}%
\left(1-s\right)-\frac{\Phi}{2\pi}\right]^{2}}{\rho_{a}^{2}}-\frac{%
\nu\lambda^{2}}{\rho_{a}^{2}}-\frac{1}{8m\rho_{a}^{2}},  \label{24}
\end{eqnarray}
where $m$ is the rest mass of the spin-half neutral particle and the
remaining terms of the energy levels (\ref{24}) correspond to the energy
levels of an one-dimensional quantum ring with radius $\rho_{a}$. We can
also see that we have an analogous expression of that one given in \cite{whw}%
, where the energy levels for the bound states depend on the geometric phase
obtained by Wei \textit{et al} \cite{whw}: $\phi_{\mathrm{WHW}%
}=2\pi\alpha\lambda B_{0}$, where $\alpha$ corresponds to the
polarizability. Note that the last term of the expression (\ref{24})
corresponds to the dynamics of a particle in a two-dimensional surface
inside a three-dimensional space as showed in \cite{costa}.

\section{Conclusions}

In this work, we have proposed a theoretical approach to study a
relativistic quantum dynamics of a spin-half neutral particle based on the
Lorentz symmetry violation background, where the wave function of the
spin-half neutral particle acquires a relativistic Abelian quantum phase
analogous to the geometric phase obtained by Wei \textit{et al} \cite{whw}
for a spinless neutral particle with an induced electric dipole moment in
the rest frame of the observers. This approach consists in the introduction
of a new term in the Dirac equation in such a way that we can obtain, in the
nonrelativistic limit of the Dirac equation, an equation of motion analogous
to that one obtained by Wei \textit{et al.} \cite{whw} in the study of the
geometric phase for a spinless neutral particle with induced electric
dipole. We have seen that this approach provides us a relativistic analogue
of the Abelian geometric quantum phase for a neutral particle with an
induced electric dipole moment, where this relativistic quantum phase is
produced by the interaction between a fixed time-like 4-vector background
and crossed electric and magnetic fields. Moreover, we have seen that this
geometric quantum phase is nondispersive. Furthermore, this approach allows
us to obtain the relativistic bound states for a spin-half neutral particle
when the neutral particle is confined to moving in a region between two
coaxial cylinder shells, showing the dependence of the energy levels on the
relativistic geometric phase produced by the Lorentz symmetry violation
background. Finally, in the nonrelativistic limit of the energy levels, we
have seen that we can obtain a dependence of the energy levels on the
geometric quantum phase analogous to that one given in Ref. \cite{whw} for a
spinless neutral particle with induced electric dipole. We would like to
comment that it is hard to discuss this relativistic approach in the
phenomenological context for a spin-half particle as done by Wei \textit{et
al.} in \cite{whw} for an induced electric dipole since we can expect that
the phase shift provided by the Lorentz symmetry violation background is
quite small, but we hope that this theoretical approach can provide new
discussions about geometric quantum phases for neutral particles, for
instance, if it can be extended for discussions about the quantum phase
obtained by Spavieri \cite{spa,spa1,spa2,spa3} for neutral particle with
electric dipole moment, and the influence of the Lorentz symmetry breaking
on the Landau quantization for neutral particles \cite{fur3}.

We would like the thank the Brazilian agency CNPq (Conselho Nacional de
Desenvolvimento Científico - Brazil) for financial support.

\end{document}